\documentclass[12pt]{article}
\usepackage{authblk}
\usepackage{graphicx}
\usepackage{amssymb}
\usepackage{amsmath}
\title{
	Analog Schwarzschild-like geometry in fluids with external pressure}

\author[]{Neven Bili\'c\thanks{bilic@irb.hr}\hspace{2pt}}
\affil[]{Theoretical Physics Division, Rudjer Bo\v skovi\'c Institute, \\
	10002 Zagreb, Croatia}

\date{\today}
\begin{document}
	\maketitle

\begin{abstract}

We study acoustic geometry in fluids with external pressure.
In particular, we examine the conditions under which the acoustic metric mimics 
a Schwarzschild-like metric. We demonstrate that  it is possible to mimic 
a Schwarzschild-like geometry in a consistent way
only in the framework of relativistic acoustic geometry.

\end{abstract}

%

\maketitle

\section{Introduction}

Analog gravity is based on the dynamics of acoustic perturbations in a fluid with 
a nonhomogeneous flow. 
Since Unruh's discovery  \cite{unruh} that a supersonic flow may cause analog Hawking radiation 
a lot of research on analog gravity has been done (see, e.g., Refs.\ \cite{barcelo2,novello} for reviews).

The acoustic perturbations in a fluid propagate in an effective curved spacetime geometry
 with the analog metric dubbed the acoustic metric.
In general, the  acoustic metric $G_{\mu\nu}$ in its relativistic form is given by \cite{moncrief,bilic,visser2}
\begin{equation}
	G_{\mu\nu} = \frac{n}{m^2 c_{\rm s} w} [g_{\mu\nu}-(1-c_{\rm s}^2)u_\mu u_\nu],
	\label{eq100}
\end{equation}
where $u_{\mu}$ is the 4-velocity of the  fluid flow in the background metric 
$g_{\mu\nu}$, usually taken to be the flat Minkowski metric,  $c_{\rm s}$ is the speed of sound in units of the speed of light,
 $n$ is the particle number density, $w$ is the relativistic specific enthalpy, and the mass scale $m$ is introduced to make the conformal factor dimensionless.
The fluid is assumed to satisfy the Euler equation with no external pressure and isentropy and  particle number conservation are usually
assumed.
In the nonrelativistic (NR) approach the acoustic metric takes the form \cite{unruh,barcelo2,visser}
 \begin{equation}
 	G_{\mu\nu}^{\rm NR} =
 	\frac{\rho_{\rm NR}}{v_{\rm s}}
 	\left(\begin{array}{ccc}
 		v_{\rm s}^2-v^2	    &   &  v_j   \\
 		&   &          \\
 		v_i &  & -\delta_{ij} 
 	\end{array} \right),
 	\label{eq118}
 \end{equation}
where $v$ is the velocity of the flow, $v_{\rm s}\equiv c_{\rm s} c$ is the speed of sound, and $\rho_{\rm NR}$ is the mass density.
The conformal factor in (\ref{eq118}) as it stands is not dimensionless. This fact is usually ignored because 
the analog spacetime line element will have the right dimension if
the metric tensor is  multiplied by a constant conformal factor of dimension $L^4 M^{-1}T^{-1}$.

In the context of acoustic geometry, a natural question arises whether it is possible to mimic the Schwarzschild spacetime, 
or even more generally, a static spacetime of the Schwarzschild-like type. 
The answer to this question is of considerable scientific importance because 
Schwarzschild-like metrics includes several  metrics of theoretical and phenomenological interest, e.g.,
the Reissner-Nordstr\"om spacetime or 
the de Sitter and anti-de Sitter spacetimes in static coordinates.
 
A few attempts were made to mimic the exact form of the Schwarzschild black hole. However, it seems that 
the corresponding acoustic metric of the form (\ref{eq100})
or (\ref{eq118}), satisfying the assumptions mentioned above, is not to be found.
As noted in Refs.\ \cite{barcelo2,novello}, assuming a constant speed of sound, it is possible to construct  a 
NR metric of the form (\ref{eq118}) 
differing from the  Schwarzschild metric
by a nonconstant conformal factor.
With nonisentropic fluids recently studied in Ref.\ \cite{bil-nik}, one has more flexibility.
However, in a recent paper \cite{bil-nik2} it was shown that the Schwarzschild metric could be reproduced in a nonisentropic fluid
only in the stiff-fluid limit when the sound speed approaches unity.
de Oliveira {\em et al.}\ \cite{deoliveira} have recently obtained a closed-form expression for an analog Schwarzschild metric in NR setup with 
a nonconstant speed of sound and a coordinate dependent external force. 
As we will shortly see, the external force can be equivalently described by a nonvanishing gradient of external pressure.
Unfortunately,  as we will demonstrate here,  the suggested scheme is not consistent,  even assuming the presence of 
an external potential
with a  nonvanishing gradient.

This paper is dealing basically with two issues: first, we derive a correct form of the acoustic metric in a fluid with external pressure,
 and second, we explore the possibility to mimic the exact  Schwarzschild black hole in such a fluid.
Effects of external pressure, or equivalently  
external potential,
have  usually been ignored in the analog gravity literature as it is claimed that upon linearization, 
the fluctuations are insensitive to any external force \cite{barcelo2}.
Here we will show that external pressure with a nonvanishing gradient may  exhibit nontrivial effects 
in the framework of both relativistic and nonrelativistic acoustic geometry.  
As to the second issue, we will describe a  procedure by which the relativistic acoustic metric exactly reproduces a Schwarzschild-like metric. To this end, we will study a spherically symmetric acoustic metric with a nonvanishing external pressure gradient. We will demonstrate that this procedure will be consistent only if the fluid is essentially relativistic.

{\bf Notation}: We use the $+---$ signature convention and, as a rule, we work in units $c=\hbar=1$. The exception is the NR regime in sections
\ref{nonrelativistic} and \ref{nonrelativistic2} where we reinstate the speed of light $c$.

The remainder of the paper is organized as follows. In section \ref{external} we discuss the relativistic
 hydrodynamics of the fluid with external pressure and derive both the relativistic and the nonrelativistic acoustic metrics.
Section \ref{schwarzschild} is devoted to the construction of 
 the Schwarzschild-like  geometry as an analog gravity model in a fluid with external pressure.  In section  \ref{conclude}, we summarize our results and  give concluding remarks. 
 
\section{Isentropic flow with external pressure}
\label{external}

 The analog acoustic geometry has been derived under the strict requirements of energy-momentum conservation,
 particle number conservation,  and 
  vanishing vorticity. 
The first two restrictions are sufficient conditions for adiabaticity.
If a stronger restriction of isentropy is assumed together with vanishing vorticity, 
the velocity field may be expressed as a potential flow, i.e., 
\cite{landau}
\begin{equation}
	w u_\mu =\partial_\mu\theta ,
	\label{eq403}
\end{equation}
where $\theta$ is the velocity potential and $w$ is the relativistic specific enthalpy. 
The reverse of the above statement is not true: a potential flow alone implies
only vanishing vorticity and 
implies neither particle number conservation nor isentropy.

Next, we  consider an isentropic fluid flow under the influence of an external  pressure field. 
We will show that if the 
external pressure, or equivalently external potential,
 has a nonvanishing gradient, 
the potential flow equation (\ref{eq403})
needs to be modified.

\subsection{Energy-momentum conservation and Euler equation}
\label{euler}

The most fundamental  assumption is the energy-momentum conservation 
\begin{equation}
 {T^{\mu\nu}}_{;\nu}=0,
  \label{eq402}
 \end{equation}
where we assume that $T^{\mu\nu}$ is  
the energy-momentum tensor of an ideal relativistic fluid 
modified to account for the presence of a conservative external force.
To this end we add 
a  term of the form $p^{\rm ext} g_{\mu\nu}$, i.e., we assume
\begin{equation}
T_{\mu\nu}=(p+\rho) u_{\mu}u_{\nu}-p g_{\mu\nu} -p^{\rm ext} g_{\mu\nu}
\label{eq019}
\end{equation}
where $p$ and $\rho$ are the fluid pressure and energy density, respectively,
$p^{\rm ext}$ is the external pressure, 
and $g_{\mu\nu}$ is the background metric.
The contraction of (\ref{eq402}) with $u_{\mu}$ 
gives 
\begin{equation}
u^\mu\rho_{,\mu}+(p+\rho){u^\mu}_{;\mu} - u^\mu p^{\rm ext}_{,\mu}=0.
 \label{eq441}
\end{equation}
Inserting this into (\ref{eq402}) with (\ref{eq019}) gives a modified  relativistic Euler equation
\begin{equation}
(p+\rho)u^{\nu}u_{\mu;\nu}
-(p+p^{\rm ext})_{,\mu}
+u^{\nu} (p+p^{\rm ext})_{,\nu}u_{\mu}  =0,
\label{eq003}
\end{equation}
which differs from the standard expression \cite{landau} by two additional terms due to external pressure.

	The assumption that an external force modifies the energy-momentum tensor
	in the form  (\ref{eq019}) may be justified by two arguments.
	The first one is based on the Lagrangian formulation. 
	Suppose that  the perfect fluid stress tensor 
	corresponds to a Lagrangian $\mathcal{L}(\varphi,X)$ that depends on the field $\varphi$
	and its kinetic term $X=g_{\mu\nu}\varphi_{,\mu} \varphi_{,\nu}$. Then, 
	$T_{\mu\nu}$ 
	is obtained by taking the functional derivative of the action with respect to $g^{\mu\nu}$, as usual. Now,  one can add to the action an external potential $U$, i.e., 
	$S=\int d^4x \sqrt{-g}(\mathcal{L}-U)$, where $U=U(x)$ is just a function of 
	$x$ (not a dynamical field). This function serves as a potential for an external force. 
	The  functional derivative of $S$  with respect to $g^{\mu\nu}$ will produce $T_{\mu\nu}$ in the form (5) with 
	$p ={\cal L}$,
	$\rho = 2 X {\cal L}_{X}-{\cal L}$, 
	$ u_\mu=\varphi_{,\mu}/\sqrt{X}$ and $p^{\rm ext}=-U$.
	
	The second argument is based on the nonrelativistic limit of the Euler
	equation. In that limit, as we will shortly demonstrate,  the energy-momentum conservation  (\ref{eq402})  with (\ref{eq019}) and the Euler equation (\ref{eq003}) in turn,  
	 yields the  
	  Euler equation in the standard nonrelativistic form in which  $p^{\rm ext}$ 
	 appears in the combination $p+p^{\rm ext}$ and hence, apparently plays the role of external pressure. Then, the external force is proportional to the gradient of $p^{\rm ext}$.

Next,  we introduce a few useful thermodynamic relations.
For a general 
thermodynamic system at nonzero temperature $T$ the first law of thermodynamics 
may be written as
\begin{equation}
dp=ndw -nTds ,
\label{eq500}
\end{equation}
where $n$ is the particle number density,  $s$ is the specific entropy, i.e., the entropy per particle, 
and $w$ is the relativistic specific enthalpy defined as
\begin{equation}
w=\frac{p+\rho}{n} .
 \label{eq404}
\end{equation}

Next, we assume that the flow is isentropic, i.e., we set $ds=0$.
For an isentropic flow, from  Eqs.\ (\ref{eq500}) and (\ref{eq404}) it follows that
\begin{equation}
w_{,\mu}=\frac{p_{,\mu}}{n},
\label{eq4018}
\end{equation}
and
\begin{equation}
	n_{,\mu}=\frac{\rho_{,\mu}}{w}.
	\label{eq4019}
\end{equation}
Using (\ref{eq404}) and (\ref{eq4018}) from Eq.~(\ref{eq003}) it follows  that
\begin{equation}\label{rnif8}
	u^{\nu}(wu_{\mu})_{;\nu}-w_{,\mu}=
	\frac{1}{n}(p^{\rm ext}_{,\mu}-u^\nu p^{\rm ext}_{,\nu} u_\mu) .
\end{equation}
Similarly,  
Eq.~(\ref{eq441}) with (\ref{eq404}) and (\ref{eq4019}) gives
\begin{equation}
	(nu^{\mu})_{;\mu}-\frac{1}{w}u^\mu p^{\rm ext}_{,\mu}=0.
	\label{eq442}
\end{equation}

Clearly,
the usual continuity equation
\begin{equation}
	(nu^{\mu})_{;\mu}=0
	\label{eq0442}
\end{equation}
 will not hold
if the  external pressure gradient is nonzero and hence,
the particle number is generally not conserved.
The particles are locally created or destroyed depending on the sign of 
$u^\mu p^{\rm ext}_{,\mu}$.
Suppose that $w$ is positive 
which, given the definition (\ref{eq404}), holds for normal matter satisfying the weak energy condition. 
Then, Eq.~(\ref{eq442}) states that the particles are locally created/destroyed if the gradient of the external pressure 
$u^\mu p^{\rm ext}_{,\mu}$ is positive/negative.
For matter that violates weak energy condition--the so called ``phantom matter''--the condition is reversed: 
the particles will be locally created/destroyed if the gradient of the  external pressure is negative/positive. 

This situation is similar to the case of a nonisentropic fluid where, as demonstrated in Ref.\ \cite{bil-nik}, 
the particle number conservation is violated for a large class of nonisentropic fluids with nonvanishing entropy gradients. 

{\bf Nota bene}: 
It may easily be demonstrated (see section \ref{nonrelativistic})
that in the NR  limit $c \rightarrow \infty$
the second term in (\ref{eq442}) is $1/c^2$ suppressed 
compared with the first term. Hence, the above-mentioned particle production is purely   a relativistic effect. 
It is worth mentioning that in most applications of thermodynamics and fluid dynamics in cosmology 
the conservation of both particle number and entropy has been assumed (see, e.g., \cite{saridakis} and references therein).

In the absence of external pressure, some of the equations would further simplify if  the enthalpy flow 
$wu_{\mu}$ were a gradient of a scalar potential, i.e., if
the velocity field satisfied Eq.\ (\ref{eq403}).
Then, as a consequence of (\ref{eq403}),
the left-hand side of (\ref{rnif8}) would vanish identically.
The assumption (\ref{eq403}) is automatically satisfied in the field-theory formalism as demonstrated in Ref.\ \cite{bil-nik}.

Obviously, equation 
(\ref{eq403}) would not solve  (\ref{rnif8}) if there existed 
external pressure
 with a nonvanishing
gradient. However,   in this case, we show that a suitably
modified potential-flow equation can solve Eq.\ (\ref{rnif8}).

\subsection{Modified potential flow}
\label{potential}

For a stationary flow with external pressure it is possible 
to define  an effective enthalpy function $W$,
\begin{equation}
	W=w+\tilde{w} ,
	\label{eq2012}
\end{equation}
 where the time independent function $\tilde{w}$  satisfies 
\begin{equation}
	u^{\nu}(\tilde{w} u_{\mu})_{;\nu}-\tilde{w}_{,\mu}
	+\frac{1}{n}(p^{\rm ext}_{,\mu}-u^\nu p^{\rm ext}_{,\nu} u_\mu)=0.
	\label{eq0008}
\end{equation}
Then, Eq.\ (\ref{rnif8}) takes the form
\begin{equation}
	u^{\nu}(W u_{\mu})_{;\nu}-W_{,\mu}
=0.
	\label{eq0007}
\end{equation}
This equation will be satisfied if the effective enthalpy flow $Wu_\mu$
is a gradient of a potential as in (\ref{eq403}), i.e., if there exist a scalar function $\theta$
such that 
\begin{equation}
	W u_\mu =\partial_\mu\theta .
	\label{eq0012}
\end{equation}

To find $\tilde{w}$
 we multiply Eq.\ (\ref{eq0008}) by the time-translation Killing vector $\xi^\mu$ and use the fact that
 $\xi^\mu p^{\rm ext}_{,\mu} = \xi^\mu \tilde{w}_{,\mu}=0$ since $p^{\rm ext}$ and $\tilde{w}$ for a stationary flow do not depend on time. 
 By making use of the Killing equation 
$\xi_{\mu;\nu}+\xi_{\nu;\mu}=0$ we find
\begin{equation}
	u^{\nu}\left[(\tilde{w} \xi^\mu u_{\mu})_{,\nu}-
\frac{1}{n} \xi^\mu u_\mu  p^{\rm ext}_{,\nu}\right]=0
	\label{eq0009}
\end{equation}
Since $u^\mu$ is basically an arbitrary timelike field we 
obtain a simple differential equation for $\tilde{w}$ 
\begin{equation}
	(\tilde{w} \xi^\mu u_{\mu})_{,\nu}=
	\frac{1}{n} \xi^\mu u_\mu  p^{\rm ext}_{,\nu}
	\label{eq0010}
\end{equation}
with solution
\begin{equation}
	\tilde{w} = \frac{1}{\xi^\mu u_{\mu}}\int
	\frac{\xi^\mu u_{\mu}}{n} d p^{\rm ext}    .
	\label{eq0011}
\end{equation} 
This integral may be thought of as a line integral along an arbitrary curve starting from a  fixed spacetime point and ending at $x$.
The integration curve  may be conveniently chosen depending on the symmetry of the flow. 
For example, for a radial flow in a static spherically symmetric spacetime
with $\xi^\mu=(1;\vec{0})$ we have 
\begin{equation}
	\tilde{w}(r)-\tilde{w}(r_0) = \frac{1}{\gamma \sqrt{g_{00}}}\int_{r_0}^r
	\frac{\gamma \sqrt{g_{00}}}{n}\partial_r p^{\rm ext} dr ,
	\label{eq0015}
\end{equation}
where $r_0$ is arbitrary, 
$\gamma= (1-v^2)^{-1/2}$ is the usual relativistic factor, and
$v$ is the radial velocity of the fluid.

Next, we derive the acoustic metric assuming a potential flow 
defined  by (\ref{eq0012}).

\subsection{Acoustic metric}
\label{acoustic}

The acoustic metric is the effective metric 
perceived by acoustic perturbations propagating in a perfect fluid
background. 
Under certain conditions,
the perturbations satisfy the Klein-Gordon equation in curved geometry 
with metric of the form (\ref{eq100}).

We first derive a propagation equation for linear perturbations 
of an isentropic flow assuming a fixed background geometry.
Given  some average bulk motion represented by
$p$, $n$, and
$u^{\mu}$, and external pressure $p^{\rm ext}$, following the standard procedure \cite{bilic}, 
we replace
\begin{equation}
	p\rightarrow p+\delta p, \quad n\rightarrow n+\delta n ,
	\quad W\rightarrow W+\delta W,
	\quad
	u^{\mu}\rightarrow u^{\mu}+\delta u^{\mu},
\label{eq108}
\end{equation}
where  
$\delta p$, 
$\delta n$, $\delta W$, and 
$\delta u^{\mu}$ 
are small disturbances.
As we do not include the  nonadiabatic perturbations, since for an isentropic flow $\delta s=0$,  we may assume that there exists an equation of state
$n=n(W)$. Using this and  
 the replacements (\ref{eq108})  in
equation
(\ref{eq442}) 
 at linear order we find 
\begin{equation}
	\left[\frac{\partial n}{\partial W}\delta W u^\mu
	+n \delta u^\mu\right]_{;\mu}=  -\frac{1}{w^2}u^\mu p^{\rm ext}_{,\mu}
	( \delta W-\delta \tilde{w})
	+\frac{1}{w} p^{\rm ext}_{,\mu}\delta u^\mu 
	\label{eq408}
\end{equation}
Here and from here on, it is understood that the partial derivatives are taken at fixed $s$.
The adiabatic perturbations $\delta W$, $\delta u^\mu$, and $\delta\tilde{w}$
may be expressed in terms of the velocity-potential perturbation $\delta\theta$.
From 
(\ref{eq0012}) 
it follows that
\begin{equation}
	\delta W=u^\mu\delta\theta_{,\mu},
	\label{eq406}
\end{equation}
\begin{equation}
	W\delta u^\mu=(g^{\mu\nu}-u^\mu u^\nu)\delta\theta_{,\nu} 
	\label{eq407}
\end{equation}
According to (\ref{eq0011}), the perturbation of the quantity $\tilde{w}$ 
is induced by the perturbation of $n$ and hence
\begin{equation}
\delta \tilde{w}=\frac{\partial\tilde{w}}{\partial n} 
\frac{\partial n}{\partial W}u^\nu\delta\theta_{,\nu} .
	\label{eq415}
\end{equation}

To simplify the notation, in the following we denote by $\chi$ the perturbation
$\delta\theta \equiv \chi$.
Besides, we introduce an effective speed of sound $\tilde{c}_{\rm s}$
defined by
\begin{equation}
	\tilde{c}_{\rm s}^{2}=
	\frac{n}{W} \frac{\partial W}{\partial n}
	=\frac{1}{1+\tilde{w}/w} \left( c_{\rm s}^2+\frac{n}{w}
	\frac{\partial\tilde{w}}{\partial n}\right),
	\label{eq0014}
\end{equation}
where $c_{\rm s}$ is the usual adiabatic speed of sound defined by
\begin{equation}
	c_{\rm s}^{2}= \frac{\partial p}{\partial \rho} =
	\frac{n}{w}\frac{\partial w}{\partial n} .
	\label{eq0016}
\end{equation}
Then, combined with (\ref{eq406})-(\ref{eq0014}), equation (\ref{eq408}) takes the form
\begin{eqnarray}
	\left(f^{\mu\nu}	\chi_{,\nu} \right)_{;\mu} = 
	\frac{1}{wW}	\left[
g^{\mu\nu}
	-\left(1+\frac{W}{w}-\frac{n}{w}\frac{\partial\tilde{w}}{\partial n} 
	\frac{1}{\tilde{c}_{\rm s}^2}\right)u^\mu u^\nu \right] 
	p^{\rm ext}_{,\mu}\chi_{,\nu} 
	\label{eq413}
\end{eqnarray}
where 
\begin{equation}
	f^{\mu\nu}=\frac{n}{W}\left[g^{\mu\nu} -\left(1-\frac{1}{\tilde{c}_{\rm s}^2}\right)u^\mu u^\nu\right] .
	\label{eq423}
\end{equation}
Applying the standard procedure  
we can recast (\ref{eq413}) into the form
\begin{equation}
	\frac{1}{\sqrt{-G}}
	\partial_{\mu}
	\left(
	{\sqrt{-G}}\,G^{\mu\nu}
	\partial_{\nu}\chi\right)
	-\frac{m^2\tilde{c}_{\rm s}W}{n^2 w}
\left[
g^{\mu\nu}
-\left(1+
\frac{c_{\rm s}^2}{\tilde{c}_{\rm s}^2}\right)u^\mu u^\nu \right] 
p^{\rm ext}_{,\mu}\chi_{,\nu} 
	=0.
	\label{eq5}
\end{equation}
Here, the matrix
\begin{equation}
	G^{\mu\nu}=\frac{1}{\omega}
	[g^{\mu\nu}-(1-\frac{1}{\tilde{c}_{\rm s}^2})u^\mu u^\nu] ,
	\label{eq2208}
\end{equation}
is the inverse of
the effective metric tensor 
\begin{equation}
	G_{\mu\nu}=\omega
	[g_{\mu\nu}-(1-\tilde{c}_{\rm s}^2)u_\mu u_\nu]\, ,
	\label{eq0013}
\end{equation}
with
\begin{equation}
	\omega=\frac{n}{m^2 \tilde{c}_{\rm s} W}
	\label{eq1013}
\end{equation}
and the determinant
\begin{equation}
	G\equiv \det G_{\mu\nu} =\omega^4 \tilde{c}_{\rm s}^2\det g_{\mu\nu} .
	\label{eq421}
\end{equation}
The mass parameter $m$
in (\ref{eq1013})
is introduced to make $G_{\mu\nu}$ dimensionless.  
Hence, the analog metric, in contrast to the standard expression (\ref{eq100}),  
involves  the effective specific enthalpy, the effective speed of sound $\tilde{c}_{\rm s}$, 
and a derivative coupling with the external pressure.

Consider next a stationary flow and time-independent external pressure.  Because of Eq.\ (\ref{eq0012}),
a stationary flow restricts the time dependence of $\theta$ so that
\begin{equation}
	\theta = mt + g(x)
	\label{eq1012}
\end{equation}
where $g$ is an arbitrary time independent function and $m$ is a constant 
which may be identified with the mass scale parameter in (\ref{eq1013}).
Equation (\ref{eq1012} together with (\ref{eq0012}) fixes the effective specific
enthalpy 
\begin{equation}
	W = \frac{m}{u_0} = \frac{m}{\gamma}.
	\label{eq1014}
\end{equation}

Next, we derive the  acoustic metric for a
NR fluid with the Newtonian gravitational field and  
an external pressure field acting on the fluid.

\subsection{Nonrelativistic limit}
\label{nonrelativistic}

The transition to NR regimes is achieved by
transforming the thermodynamic functions as follows:
\begin{equation}
	n\rightarrow \frac{\rho_{\rm NR}}{m},
	\quad \rho \rightarrow \rho_{\rm NR} c^2+\varepsilon, \quad
	w\rightarrow mc^2 + m h ,
	\label{eq011}
\end{equation}
where $\rho_{\rm NR}$,   $\varepsilon$, and $h$ are the NR mass, energy,
and enthalpy  densities, respectively.
Next, the NR versions of the  continuity 
and Euler
equations are obtained  by making use of the background 
  metric in the post-Newtonian gauge \cite{will,blanchet,poisson} 
\begin{equation}
	ds^2=\left(1+2\frac{\Phi}{c^2}\right) (cdt)^2-\left(1-2\frac{\Phi}{c^2}\right)
	d\vec{x}^2 ,
\end{equation}
where $\Phi=\Phi(\vec{x})$ is a Newtonian-like 
potential.\footnote{Note that we are using the Landau-Lifshitz convention for the sign of the Newton potential, i.e., \\ $\Phi(r)=-GM/r$.}
Besides, we make use of the prescription $v_i \rightarrow v_i/c$ and 
\begin{equation}
\gamma=\left(1-\frac{v^2}{c^2}\right)^{-1/2}, \quad
	u^\mu =\gamma (1/\sqrt{g_{00}}\, ,\vec{v}/c),  \quad u_\mu =\gamma (\sqrt{g_{00}}, -\vec{v}/c), 
	\quad \partial_\mu = \left(\frac{\partial}{c\partial t}, \nabla\right).
\label{eq6005}
\end{equation}
Then, applying  this  to (\ref{eq003})  and keeping 
the leading terms in $1/c^2$ we  find the NR Euler equation
\begin{equation}
	\rho_{\rm NR} 
	\left( \frac{\partial \vec{v}}{\partial t}+ \frac12 \nabla v^2
	-	\vec{v}\times\nabla \times\vec{v}\right)
	+\rho_{\rm NR}\nabla \Phi 
	+\nabla(p+p^{\rm ext}) =0.
	\label{eq005}
\end{equation} 
Assuming, as usual,  that the fluid is  irrotational,
the velocity may be written as a gradient of a scalar potential, i.e.,
\begin{equation}
	\vec v = -\nabla \theta ,
	\label{eq006}
\end{equation}
which is a NR version of (\ref{eq403}).
In this case Eq.\ (\ref{eq005}) simplifies to
\begin{equation}
	\nabla 
	\left( -\frac{\partial \theta}{\partial t}+ \frac12 (\nabla \theta )^2
	\right) 
+\nabla \Phi
+\frac{1}{\rho_{\rm NR}}\nabla(p+p^{\rm ext}) =0,
\label{eq007}
\end{equation} 
which may be  reduced to the Bernoulli equation 
\begin{equation}
	-\frac{\partial \theta}{\partial t}+ \frac12  v^2
+\Phi
 +h + V = {\rm const},
	\label{eq008}
\end{equation}
where we have introduced the NR specific enthalpy $h$ and the external potential
$V$
\begin{equation}
	\nabla h = \frac{1}{\rho_{\rm NR}} \nabla p , 
	\label{eq009}
\end{equation}
\begin{equation}
	\nabla V=	\frac{1}{\rho_{\rm NR}} \nabla p^{\rm ext} .
	\label{eq010}
\end{equation}
 Clearly, the quantity $V$ is a potential for
an external force $\vec{F} = \nabla V$ acting on the fluid.

Equations (\ref{eq009}) and (\ref{eq010}) are the NR versions of
equations  (\ref{eq4018}) and (\ref{eq0010}), respectively.

Next, applying  (\ref{eq6005})  to either (\ref{eq441}) or (\ref{eq442}) and keeping 
the leading and next to leading terms in $v/c$ expansion,  we  find a modified NR continuity equation
\begin{equation}
	\partial_t \rho_{\rm NR} +\nabla (\rho_{\rm NR}\vec{v})=
\frac{1}{c^2} \vec{v}\,\nabla(p+2 p^{\rm ext})+\frac{3}{c^2}
 \rho_{\rm NR}\vec{v}\,\nabla \Phi +\mathcal{O}(v^4/c^4).
	\label{eq603}
\end{equation}
Clearly, the right-hand side is a relativistic correction and 
hence, in the limit $c\rightarrow \infty$, the NR continuity equation 
\begin{equation}
	\partial_t \rho_{\rm NR} +\nabla (\rho_{\rm NR}\vec{v})= 0
\label{eq1603}
\end{equation}
holds.  

 Now we proceed to derive the NR acoustic metric for a fluid subjected to external 
potential.
    Following Visser \cite{visser}, we start from 
  the potential flow equation (\ref{eq006}), Bernoulli equation (\ref{eq008}), and
the NR version of the continuity equation (\ref{eq1603}).
 We linearize  Eqs.\ (\ref{eq008}) and 
(\ref{eq1603}) by  replacing
\begin{equation}
\rho_{\rm NR} \rightarrow \rho_{\rm NR}+ \delta \rho_{\rm NR}, \quad
h \rightarrow h+\delta h, \quad 
\theta \rightarrow \theta+\delta \theta,
\end{equation}
where the quantities $\delta \rho_{\rm NR}$, $\delta h$, and $\delta \theta$
are small acoustic perturbations around some average bulk motion represented by
 $\rho_{\rm NR}$, $ h$, and $\theta$. For notation simplicity, in the following equations we shall use $\phi$ instead of $\delta\theta$, keeping in mind that 
 $\phi$ is infinitesimally small. Then, Eqs.\ (\ref{eq008}) and  (\ref{eq1603}) 
 at linear order give
 \begin{equation}
 	\partial_t \phi + \vec{v}\nabla \phi -\delta h -\delta V=0 ,
 \label{604}
 \end{equation} 
and 
\begin{equation}
	\partial_t \delta\rho_{\rm NR} + \nabla(\vec{v}\delta\rho_{\rm NR}  
-\rho_{\rm NR} \nabla\phi )	=0 .
	\label{605}
\end{equation}
Owing to Eqs.\ (\ref{eq009}) and (\ref{eq010}) the specific enthalpies $h$ and
$V$ may be regarded as implicit functions of $\rho_{\rm NR}$. Then, by making use of  Eq. (\ref{604}) and mathematical identity 
\begin{equation}
	\delta h +\delta V=\left( \frac{\partial h}{\partial\rho_{\rm NR}} +\frac{\partial V}{\partial\rho_{\rm NR}}\right)\delta\rho_{\rm NR},
\label{eq606}
\end{equation}
the variation $\delta\rho_{\rm NR}$ can expressed as
\begin{equation}
\delta\rho_{\rm NR}=\left( \frac{\partial h}{\partial\rho_{\rm NR}} +\frac{\partial V}{\partial\rho_{\rm NR}}\right)^{-1}
(\partial_t \phi + \vec{v}\,\nabla \phi).
	\label{eq607}
\end{equation}
Substituting this into (\ref{605}) we find 
\begin{equation}
	\partial_t \left[\frac{\rho_{\rm NR}}{\tilde{v}_{\rm s}^2}(\partial_t \phi + \vec{v}\,\nabla \phi)\right]
+\nabla\left[\frac{\rho_{\rm NR}}{\tilde{v}_{\rm s}^2}(\partial_t \phi + \vec{v}\,\nabla \phi)\vec{v} -\rho_{\rm NR}\nabla\phi  \right]=0 ,
	\label{608}
\end{equation}
where we have introduced an effective speed of sound squared  
\begin{equation}
\tilde{v}_{\rm s}^2=\left( \frac{\partial h}{\partial\rho_{\rm NR}} +\frac{\partial V}{\partial\rho_{\rm NR}}\right)\rho_{\rm NR}
=v_{\rm s}^2+\frac{\partial V}{\partial\rho_{\rm NR}}\rho_{\rm NR}.
	\label{eq609}
\end{equation}
Here,   $v_{\rm s}$  is the usual adiabatic  speed of sound
defined by
\begin{equation}
	v_{\rm s}^2=\left. \frac{\partial p}{\partial\rho_{\rm NR}}\right|_s .
\end{equation}

 Equation  (\ref{608}) can be used to construct  the   acoustic metric in its NR form. Applying the usual procedure
\cite{visser} we find 
\begin{equation}
	G_{\mu\nu} =
\frac{\rho_{\rm NR}}{\tilde{v}_{\rm s}}
	\left(\begin{array}{ccc}
	\tilde{v}_{\rm s}^2-v^2	    &   &  v_j   \\
		&   &          \\
		v_i &  & -\delta_{ij} 
	\end{array} \right),
	\label{eq018}
\end{equation}
precisely in the form of (\ref{eq118}) except that the standard speed of sound $v_{\rm s}^2$ is replaced by
$\tilde{v}_{\rm s}^2$. As mentioned in Introduction,  the conformal factor can be made dimensionless by multiplying it by a suitable dimensionful factor.
In the following, we will assume that this factor is absorbed in 
the mass density $\rho_{\rm NR}$ so that the quantity $\rho_{\rm NR}$ in 
the equations of the NR acoustic geometry 
will appear to have the dimension $L/T$, i.e., the dimension of velocity.

Obviously, for a fluid without external 
potential,
the metric (\ref{eq018}) coincides with  the NR acoustic metric (\ref{eq118}).

\section{Schwarzschild-like analog geometry}
\label{schwarzschild}
As an application of  the formalism presented in section \ref{external}, 
in this section, we proceed toward the construction of a spherically symmetric analog black hole using 
the approach of de Oliveira {\em et al.}\ \cite{deoliveira}. 
In their approach, the conservation of the particle number was imposed
 and, 
 to maintain  
the Euler equation and the correct 
definition of the speed of sound, it was necessary to 
introduce an external force field, or equivalently, external pressure 
with a nonvanishing gradient. Here we demonstrate that it is not possible to simultaneously satisfy the continuity equation and the definition of the speed of sound even in the presence of external pressure with a nonvanishing gradient.
We will show that a consistent approach requires  giving up the continuity equation so that the violation of the particle number conservation  is compensated by  external pressure with a nonvanishing gradient.

\subsection{Relativistic approach}
\label{relativistic}
Consider a relativistic fluid subjected to an external pressure field. 
To mimic a Schwarzschild-like geometry in analog gravity we assume a spherical symmetry with radial three-velocity $v$.
We start from the expression 
 (\ref{eq0013}) applied to a flow in Minkowski spacetime with spherical symmetry
 and by suitable coordinate transformation $t=t(T,R), r=r(T,R)$
 we will try to mimic a spacetime with line element of the form
 \begin{equation}
 	ds^2=fdT^2-\frac{1}{f}dR^2 -R^2 d\Omega^2
 	\label{eq0017}
 \end{equation}
 where $f$ is a function of $R$ and $f=0$ at the horizon. 
 We first transform the metric (\ref{eq0013}) 
 with 
  \begin{equation}
 	g_{\mu\nu}={\rm diag}(1,-1,-r^2,-r^2\sin^2\vartheta) ,
 	\label{eq1021}
 \end{equation}
 into diagonal form using a coordinate transformation
 \begin{equation}
 	dT= dt +\frac{(1-\tilde{c}_{\rm s}^2)\gamma^2 v}{1-(1-\tilde{c}_{\rm s}^2)\gamma^2} dr,
 	\label{eq0018}
 \end{equation}
where we  use the following notation:
\begin{equation}
	u_\mu=(\gamma, -\gamma v), \quad \gamma= (1-v^2)^{-1/2}.
\label{eq0019}
\end{equation}
The transformation (\ref{eq0018}) is similar to  
the Gullstrand–Painlev\'e type of coordinate transformation used recently 
by Hossenfelder \cite{hossenfelder} to construct an analog  AdS planar black hole.
By making use of a new radial coordinate $R$ defined by
\begin{equation}
	R^2=\omega r^2,
\label{eq0020}
\end{equation}
where $\omega$ is given by (\ref{eq1013}),
we obtain the analog line element in the form
 \begin{equation}
	ds^2=\frac{R^2}{r^2} \left[(\tilde{c}_{\rm s}^2-v^2)\gamma^2 dT^2-
	\frac{\tilde{c}_{\rm s}^2}{(\tilde{c}_{\rm s}^2-v^2)\gamma^2}\frac{dR^2}{(R')^2}\right] -R^2 d\Omega^2.
	\label{eq0021}
\end{equation}
where the prime $'$ denotes a derivative with respect to $r$.
To reproduce the metric (\ref{eq0017}), we demand 
\begin{equation}
f=\frac{R^2}{r^2} (\tilde{c}_{\rm s}^2-v^2)\gamma^2 
=\frac{r^2 (R')^2(\tilde{c}_{\rm s}^2-v^2)\gamma^2}{R^2\tilde{c}_{\rm s}^2},
	\label{eq0022}
\end{equation}
as a constraint 
on $r$-dependent quantities $R$, $v$, and $\tilde{c}_{\rm s}$. 
Clearly, the acoustic horizon defined by  
 $v^2=\tilde{c}_{\rm s}^2$ coincides with the horizon $f=0$ of the Schwarzschild-like black hole.
From (\ref{eq0022}) we find $v^2$ and $\tilde{c}_{\rm s}^2$ expressed in terms of $R$ and $r$.
\begin{equation}
	v^2=\frac{\tilde{c}_{\rm s}^2-fr^2/R^2}{1-fr^2/R^2},
	 \quad	\gamma^2=\frac{1-fr^2/R^2}{1-\tilde{c}_{\rm s}^2},
	\label{eq0023}
\end{equation}
\begin{equation}
\tilde{c}_{\rm s}^2=\frac{r^4}{R^4} (R')^2
	\label{eq0024}
\end{equation}
In addition,  we have an equation which connects
$n$ and   $\tilde{c}_{\rm s}^2$ through the conformal factor $\omega$. Using the definition
(\ref{eq1013}) with (\ref{eq1014}), (\ref{eq0020}), and (\ref{eq0024}) we find
\begin{equation}
	n^2=m^6 (1-v^2) (R')^2
	\label{eq0026}
\end{equation}
The system of three equations (\ref{eq0023})-(\ref{eq0026}) is not complete since we have four variables. 
We still have two conditions that must be met: the continuity equation (\ref{eq442}) and the definition of the effective
speed of sound  (\ref{eq0014}).

Consider first the case of zero external pressure.
In this case the effective speed of sound $\tilde{c}_{\rm s}$ becomes equal to 
the standard adiabatic speed of sound $c_{\rm s}$, the second term in Eq.\ (\ref{eq442}) vanishes, 
and the continuity equation (\ref{eq0442}) applies.
This equation then
implies
\begin{equation}
	n=\frac{\alpha m}{r^2}\frac{(1-\tilde{c}_{\rm s}^2)^{1/2}}{(\tilde{c}_{\rm s}^2-fr^2/R^2)^{1/2}}, 
	\label{eq0025}
\end{equation}
where $\alpha$ is an arbitrary positive dimensionless constant.
Then, combining Eqs. (\ref{eq0023})-(\ref{eq0025})
we find a differential equation for $R$
\begin{equation}
	\frac{\alpha^2R^4}{m^4}\left(1-\frac{r^2f}{R^2}\right)-r^8(R')^4
	+r^6 f R^2(R')^2=0.
	\label{eq0027}
\end{equation}
This equation will coincide with its NR version derived in Ref.\ 
\cite{deoliveira} if we identify $\alpha/m^2$ with their scale $k$ and neglect
the second term in brackets on the left-hand side of (\ref{eq0027}). This term is 
 a relativistic correction of the order of $v^2/c^2$, as can be shown by applying  
  the NR limit to the second equality in (\ref{eq0023}).
It is important to stress that Eq.\ (\ref{eq0027}) has been derived under the  requirement of particle number conservation, i.e., assuming  
Eq.\ (\ref{eq0442}).

Next, we discuss what would happen if, instead of 
the continuity equation (\ref{eq0442}), we imposed  Eq.\ (\ref{eq0014}). Although this equation reduces to (\ref{eq0016}) 
in the case of nonvanishing external pressure, the following analysis applies to both vanishing and nonvanishing external pressure. 
In either case, we  would obtain a differential equation for  $R$
 different from (\ref{eq0027}).
Indeed, from Eqs.\ (\ref{eq0014}) with (\ref{eq1014}), (\ref{eq0024}), and (\ref{eq0026}) we find 
\begin{equation}
\frac{\gamma'}{\gamma}\left( 1- \frac{(R')^2r^4}{R^4}	\right) =-\frac{R' R'' r^4}{R^4}
	\label{eq0029}
\end{equation}
where 
\begin{equation}
\gamma=\frac{(1-fr^2/R^2)^{1/2}}{(1-r^4 (R')^2/R^4)^{1/2}} .
	\label{eq0030}
\end{equation}
Note that Eq.\ (\ref{eq0029})  with (\ref{eq0030}) holds for a fluid with or without  external pressure.
Clearly, Eqs.\ (\ref{eq0027}) and (\ref{eq0029}) are substantially different but
it is not immediately apparent that  these equations  are incompatible with each other. To make a comparison, we recast (\ref{eq0027}) 
into a form similar to (\ref{eq0029}) by making use of (\ref{eq0023}) and the logarithmic derivative. 
In this way we arrive at  a second order differential equation 
\begin{eqnarray}
	& &  \frac{\gamma'}{\gamma}\left( 1- \frac{(R')^2r^4}{R^4}\right) =  
\frac{R''}{R'}+	\frac12 (R')^2\left(\frac{r^4}{R^4}	\right)' 
	\nonumber \\
	& &+\left( 1- \frac{(R')^2r^4}{R^4}\right)
	\left[\frac{3}{r}  +\frac{R'}{R}   
	+\frac12 \frac{(r^2(R'/R)^2-f)'}{r^2(R'/R)^2-f} 
	\right]	. 
	\label{eq1029}
\end{eqnarray}
In contrast to  Eq.\ (\ref{eq0029}), the right-hand side of Eq.\ (\ref{eq1029})  involves a dependence on $f$, and hence,
 equations (\ref{eq0027}) and (\ref{eq0029}) cannot be simultaneously satisfied. 
 In other words,   
the particle number conservation expressed by the continuity equation (\ref{eq0442}) and the defining equation for the speed of
sound (\ref{eq0014}) cannot be made compatible with each other.
Hence, the scheme that involves external force suggested in Ref.\ \cite{deoliveira} seems to be inconsistent in the relativistic approach.
This inconsistency can be demonstrated also in the NR approach which we discuss next in Sec. \ref{nonrelativistic2}.
	
A consistent approach would require giving up  the particle number conservation (\ref{eq0442}) and adhering to the definition of the speed of sound.
As shown in section \ref{euler}, 
the condition of particle number conservation is altered in the presence of  external pressure with a nonvanishing gradient.
 Hence, with a suitable choice of external pressure, a modified continuity equation  (\ref{eq442})
  can be made compatible with Eq.\ (\ref{eq0014}).
 
 To find the explicit expression for the required   external pressure gradient we start from Eq.\ (\ref{eq442}) which,
assuming a stationary radial flow, takes the form 
\begin{equation}
\frac{w}{r^2}(r^2n\gamma v)'=\gamma v(p^{\rm ext})' .
	\label{eq2044}
\end{equation}
With the help of  
(\ref{eq2012}), (\ref{eq0015}), and (\ref{eq1014}),
equation (\ref{eq2044}) can be expressed as a differential equation  for the external pressure gradient $(p^{\rm ext})'$,
\begin{equation}
-\frac{\gamma}{n}(p^{\rm ext})'=\left(\frac{\gamma^2r^2 v (p^{\rm ext})'}{(\gamma r^2 v n)'}\right)',
	\label{eq1016}
\end{equation}
with solution
\begin{equation}
(p^{\rm ext})'=	\beta m^2 \frac{(\gamma r^2 v n)'}{\gamma^2r^2 v(\gamma r^2 v n)}, 
	\label{eq1018}
\end{equation}
where $\beta$ is a dimensionless integration constant.
The quantities $v$, $\gamma$, and $n$ can be expressed in terms of $r$, $R$, and $R'$ via Eqs.\ (\ref{eq0023})--(\ref{eq0026}). 
 Hence, in conjunction with Eq.\ (\ref{eq0029}) which can be solved for $R$,
 Eq.\ (\ref{eq1018}) yields the external pressure gradient as an explicit function of $r$.

Comparing Eqs.\ (\ref{eq2044}) and (\ref{eq1018}) we also obtain a closed expression for $w$,
\begin{equation}
	w=	\frac{\beta m^2 }{\gamma^2 r^2 v n}.
	\label{eq2018}
\end{equation}
Once the function $R=R(r)$ is obtained as a solution to (\ref{eq0029}), we can 
determine the equation of state of the fluid $w=w(n)$ as a parametric function
defined  by  (\ref{eq0026}) and (\ref{eq2018}).
Hence we have fixed everything: the velocity profile, the equation of state, and the external pressure gradient.

	It would be of considerable interest to study the existence and uniqueness of solutions to
Eq. (\ref{eq0029}). This equation is 
  a rather complicated nonlinear differential equation and
	it is difficult to analytically prove or disprove the existence and uniqueness of its solutions. 
	However, it would be straightforward to  look for a numerical solution on an interval of $r$, 
	similar to what has been done in Ref.\ \cite{deoliveira} for the nonrelativistic version of Eq.\  (\ref{eq0027}) 
	but this task goes beyond the scope of this paper.
	Nevertheless, even if Eq.\ (\ref{eq0029}) has  more than one solution or a multivalued function as a solution, one can  
	single out a particular branch. For example, if $R(r)$ is a solution 
	to (\ref{eq0029}), it is clear that $-R(r)$ is also a solution and we can single out the positive one.

\subsection{Nonrelativistic approach}
\label{nonrelativistic2}
We now repeat the procedure described in the previous section using the  NR form of the acoustic metric (\ref{eq018}). 
In our derivation, unlike in  Ref.\ \cite{deoliveira},  
we assume   the existence of 
an external potential,
 from the very beginning. 
We will show that it is not possible to mimic a Schwarzschild-like
metric in a NR setup.
 
For dimensional reasons we introduce an arbitrary constant $v_0$ of the dimension of velocity and write the metric (\ref{eq0017}) as 
 \begin{equation}
	ds^2=f v_0^2 dT^2-\frac{1}{f}dR^2 -R^2 d\Omega^2.
	\label{eq1017}
\end{equation}
As we are dealing with NR fluid, we assume  $v_0 \ll c$.
Assuming spherical symmetry, we transform the NR acoustic metric (\ref{eq018})
into a diagonal form using a coordinate transformation
\begin{equation}
	T= t +\int \frac{vdr}{\tilde{v}_{\rm s}^2-v^2},
	\label{eq0031}
\end{equation}
and 
\begin{equation}
	R^2=\frac{\rho_{\rm NR}}{\tilde{v}_{\rm s}}r^2.
\label{eq0032}
\end{equation}
With this, we obtain the analog line element in the form similar to 
(\ref{eq0021})
 \begin{equation}
	ds^2=\frac{R^2}{r^2} \left[(\tilde{v}_{\rm s}^2-v^2) dT^2-
	\frac{\tilde{v}_{\rm s}^2}{(\tilde{v}_{\rm s}^2-v^2)}\frac{dR^2}{(R')^2}\right] -R^2 d\Omega^2.
	\label{eq0033}
\end{equation}
As before, to reproduce the metric (\ref{eq1017}), we impose  
\begin{equation}
	f=\frac{R^2(\tilde{v}_{\rm s}^2-v^2)}{r^2v_0^2}  
	=\frac{r^2 (R')^2(\tilde{v}_{\rm s}^2-v^2)}{R^2\tilde{v}_{\rm s}^2},
	\label{eq0034}
\end{equation}
as a constraint 
on $r$-dependent quantities $R$, $v$, and $\tilde{v}_{\rm s}$. 
Then,  $\tilde{v}_{\rm s}^2$,  $v^2$, and $\rho_{\rm NR}^2$  can be expressed in terms of $r$, $R$, and $R'$: 
\begin{equation}
\frac{\tilde{v}_{\rm s}^2}{v_0^2}=\frac{r^4}{R^4} (R')^2 , 	
	\label{eq1035}
\end{equation}
\begin{equation}
\frac{v^2}{v_0^2}=\frac{r^4}{R^4} (R')^2-\frac{fr^2}{R^2},	
\label{eq0035}
\end{equation}
\begin{equation}
 \rho_{\rm NR}^2=v_0^2(R')^2 .	
	\label{eq0036}
\end{equation}

Now we 
impose the continuity equation (\ref{eq1603}). Assuming time independence 
we have
\begin{equation}
\partial_r (r^2 \rho_{\rm NR} v)=0.
\label{eq0037}
\end{equation}
This, together with (\ref{eq0036}) gives another expression for $v^2$
\begin{equation}
\frac{v^2}{v_0^2}	= \frac{k^2}{r^4 (R')^2},
\label{eq0038}
\end{equation}
where $k$ is a constant of dimension of length squared. Plugging this into Eq.\  (\ref{eq0035})
we find a differential equation 
\begin{equation}
k^2 R^4-r^8(R')^4
	+r^6 f R^2(R')^2=0 .
	\label{eq0039}
\end{equation}
This equation, previously obtained in Ref.\ 
\cite{deoliveira},
 is the NR version of (\ref{eq0027}).
 
We still must satisfy equation 
(\ref{eq609}) because it was used in the derivation of the acoustic metric.
For  a spherically symmetric radial flow we have
\begin{equation}
v_{\rm s}^2	= \frac{p'}{\rho_{\rm NR}'}, \quad\quad 
\frac{\partial V}{\partial\rho_{\rm NR}}=\frac{1}{\rho_{\rm NR}}
\frac{(p^{\rm ext})'}{\rho_{\rm NR}'},
	\label{eq0040}
\end{equation}
and, hence,
\begin{equation}
	\tilde{v}_{\rm s}^2=v_{\rm s}^2+\frac{(p^{\rm ext})'}{\rho_{\rm NR}'}=
	\frac{p'+(p^{\rm ext})'}{\rho_{\rm NR}'}. 
	\label{eq0041}
\end{equation}
Can we use this equation to  determine 
the 
 external force
required to maintain the desired 
flow as it was done in Ref.\ \cite{deoliveira}?
To answer this question we make use of
the Euler equation for a stationary radial flow
\begin{equation}
	v v' +\frac{p'+(p^{\rm ext})'}{\rho_{\rm NR}} =0.
	\label{eq0042}
\end{equation}
Combining this with Eq.\ (\ref{eq0041}) we find  
\begin{equation}
	(v^2)'=-2 \tilde{v}_{\rm s}^2\frac{\rho_{\rm NR}'}{\rho_{\rm NR}}.
	\label{eq0048}
\end{equation}
Then, using (\ref{eq1035}) and (\ref{eq0036}) we obtain
\begin{equation}
	(v^2)'=-2{v_0^2}\frac{r^4}{R^4}R'R''.
	\label{eq0043}
\end{equation}
Note that this equation does not involve a dependence on the external pressure gradient. Furthermore,
equation (\ref{eq0043}) together with Eq. (\ref{eq0035}) yields
another differential equation for $R$,
\begin{equation}
4r^3(RR'R''-(R')^3)+4r^2R(R')^2+2rfR^2R'-(2f+rf')R^3=0 ,
\label{eq0044}
\end{equation}
which holds  for a fluid with or without external pressure.
This equation can be obtained directly by way of the NR limit of Eq.\ (\ref{eq0029}) with (\ref{eq0030}). 

It may easily be verified that Eqs.\ (\ref{eq0039}) and (\ref{eq0044}) are not equivalent. These two equations are not directly comparable since 
Eq.\ (\ref{eq0039}) does not involve the second derivative of $R$ with respect to $r$.
To make a comparison, one can divide Eq.\ (\ref{eq0039})  by $R^4$ and take a derivative.
 Then, it may be seen that the equation thus obtained is substantially different from (\ref{eq0044}).
Hence, the continuity equation (\ref{eq0037}) and the defining equation for $\tilde{v}_{\rm s}$, Eq.\ (\ref{eq609}), 
cannot be simultaneously satisfied and hence, Eq.\ (\ref{eq0041}) cannot be used 
to determine the external force needed to maintain the desired flow.

As in the previous section, we must give up the continuity equation (\ref{eq0442}). 
This does not seem to present  a problem because, as shown in Sec.\ \ref{euler},
external pressure modifies
 the continuity equation anyway. However, 
  this modification  is a 
relativistic correction of the order of $v^2/c^2$ as shown by (\ref{eq603}), so
we can expect that the needed external pressure must be
 of the order of $c^2/v^2$
  to make its contribution  significant. 
  To find the  required external pressure gradient we start from Eq.\ (\ref{eq603})
for  a spherically symmetric flow
  and assume that the pressure terms on 
the right-hand side dominate the Newtonian gravity term. 
Combining (\ref{eq603})   with (\ref{eq0042}) and 
neglecting the gravity term,  we obtain
\begin{equation}
(p^{\rm ext})'=\frac{c^2}{r^2 v}(r^2\rho_{\rm NR}v)' +  \rho_{\rm NR}v v' .
\label{eq1044}
\end{equation}
One can easily verify that Eq.\. (\ref{eq1044}) coincides with 
the NR limit of (\ref{eq2044}) including the lowest relativistic correction. 
From (\ref{eq0035})  and (\ref{eq0036}) it follows that 
the first  term on  the right-hand side of this expression is of the order of
$c^2 v'$, whereas 
the second  term is of the order of $v_0^2 v'$. Hence, the first term
is dominant unless the velocity scale $v_0$  is close to the speed of light.
 Either way, the external pressure gradient is of the order 
\begin{equation}
	(p^{\rm ext})'\sim c^2 v'.
	\label{eq1045}
\end{equation}
On the other hand, according to (\ref{eq0042}), the gradients of both pressure and external pressure are of the order 
\begin{equation}
p'\sim	(p^{\rm ext})'\sim v_0^2 v' .
	\label{eq1046}
\end{equation}
Comparing  (\ref{eq1045}) and (\ref{eq1046}) we conclude that the velocity scale  $v_0\sim c$ 
and, as a consequence, the velocity of the fluid must be
of the order of $v\sim c$. Moreover, this implies that the speed of sound
 $v_{\rm s}$  will be close 
to the speed of light as can be deduced from Eqs.\ (\ref{eq1035})  
and (\ref{eq0040}).

Hence, to construct  a Schwarzschild-like
 metric  one needs  a  relativistic radial fluid flow  
in which  $v$ and $v_{\rm s}$ are both close to the speed of light.
This result  contradicts the initial assumption  that the fluid is nonrelativistic.
In this way, we have demonstrated that
it is not possible to mimic a Schwarzschild-like
metric in the framework of NR acoustic geometry.

\section{Summary and conclusions}
\label{conclude}

We have presented  a formulation of analog gravity in a fluid subjected to  external pressure,
or equivalently to external potential. 
We have shown that the external pressure with a nonvanishing gradient may  exhibit nontrivial effects 
 in both relativistic and  nonrelativistic approaches.  
Using this we have investigated the possibility to mimic the exact form of the  Schwarzschild black hole in such a fluid.
Transforming the metric to a diagonal form we have found the conditions under which the metric 
exactly reproduces a Schwarzschild-like metric. We have found that  
this procedure can be carried out consistently if the fluid is essentially relativistic.

The attempt to adapt the relativistic  procedure of Sec.\ \ref{relativistic} to  a NR fluid has failed 
because it is not possible to satisfy simultaneously 
the NR continuity equation and the definition of the speed of sound. 
 The reason behind this failure is that NR fluid dynamics neglects the relativistic effect of particle creation or destruction
caused  by an external pressure gradient.
The considerations in Sec.\ \ref{nonrelativistic2} lead to the conclusion that to mimic a Schwarzschild-like
metric  one needs  a  relativistic fluid with a radial flow  
in which  the fluid velocity $v$ and the speed of sound $v_{\rm s}$ are both close to the speed of light.
In other words,
it is impossible to mimic a  Schwarzschild-like metric by analog gravity in the framework of NR acoustic geometry, 

One may 
	 wonder whether a different definition of the external
	pressure--e.g., one corresponding to another ideal fluid rather than
	to one of the vacuum energy type as in Eq.\ (\ref{eq019})--might resolve  the 	mentioned
disagreement with Ref.\ \cite{deoliveira}.
Unfortunately, this attempt is unlikely to succeed, mainly for
the reason that,
 as demonstrated here,  the NR formalism 
 is inherently incompatible  with a Schwarzschild-like analog  metric.
The formalism   
 which employs the nonrelativistic acoustic geometry in  the form derived originally for 
a {\em{single}} ideal fluid while resorting to the use of an external force,
is not consistent.
An external force basically amounts to
adding another ideal fluid
which 
 necessarily requires a reformulation of the acoustic geometry. 
 We have implemented 
such a reformulation here for the external pressure defined by Eq.\ (\ref{eq019}).
If  we had used another  	
definition of the external pressure, e.g., by replacing $p$ by
 $p+p^{\rm ext}$ in the ideal fluid expression for $T_{\mu\nu}$,
we  would have  obtained
 the nonrelativistic limit of the Euler equation 
still in  the form (\ref{eq005}), so Eqs. (\ref{eq006})-(\ref{eq010}) 
would have remained the same.
As the definition of the speed of sound (\ref{eq0016})
involves the pressure $p$
rather than $p+p^{\rm ext}$,  an effective speed of sound as defined in (\ref{eq609})
would appear in the derivation of the nonrelativistic acoustic
metric. As a result,  the acoustic metric would inevitably have the form 
(\ref{eq018})
and the above-mentioned inconsistency problem 
would remain.

\section*{Acknowledgments}

The work of N.B.\ has been partially supported by the ICTP - SEENET-MTP Project No.\  NT-03 
Cosmology - Classical and Quantum Challenges.

\end{document}